# Quantum hybridization negative differential resistance from non-toxic halide perovskite nanowire heterojunctions and its strain control

Juho Lee[1,†], Muhammad Ejaz Khan[1,2,†], and Yong-Hoon Kim[1,*]

[1] Department of Electrical Engineering Korea Advanced Institute of Science and Technology (KAIST), 291 Daehak-ro, Yuseong-gu, Daejeon 34141, Republic of Korea

* Corresponding author: y.h.kim@kaist.ac.kr

[†] These authors contributed equally to this work.

**ABSTRACT:** While low-dimensional organometal halide perovskites are expected to open up new opportunities for a diverse range of device applications, like in their bulk counterparts, the toxicity of Pb-based halide perovskite materials is a significant concern that hinders their practical use. We recently predicted that lead triiodide ($PbI_3$) columns derived from trimethylsulfonium (TMS) lead triiodide $(CH_3)_3SPbI_3$ ($TMSPbI_3$) by stripping off TMS ligands should be semimetallic, and additionally ultrahigh negative differential resistance (NDR) can arise from the heterojunction composed of a $TMSPbI_3$ channel sandwiched by $PbI_3$ electrodes. Herein, we computationally explore whether similar material and device characteristics can be obtained from other one-dimensional halide perovskites based on non-Pb metal elements, and in doing so deepen the understanding of their mechanistic origins. First, scanning through several candidate metal halide inorganic frameworks as well as their parental form halide perovskites, we find that the germanium triiodide ($GeI_3$) column also assumes a semimetallic character by avoiding the Peierls distortion. Next, adopting the bundled nanowire $GeI_3$-$TMSGeI_3$-$GeI_3$ junction configuration, we obtain a drastically high peak current density and ultrahigh NDR at room temperature. Furthermore, the robustness and controllability of NDR signals under strain are revealed, establishing its potential for flexible electronics applications. It will be emphasized that, despite the performance metrics notably enhanced over those from the $PbI_3$-$TMSPbI_3$-$PbI_3$ case, these device characteristics still arise from the identical quantum hybridization NDR mechanism.

## 1. INTRODUCTION

Research in halide perovskites field has rapidly progressed due to their potential for optoelectronic applications such as solar cells, lasers, photodetectors, and light-emitting diodes [1-5]. However, excellent materials properties like defect tolerance, long charge carrier diffusion length, low cost, facile synthesizability, etc. make them also promising for non-optoelectronic device applications. Recently, we predicted that the one-dimensional (1D) inorganic lead triiodide ($PbI_3$) framework derived from the trimethylsulfonium lead triiodide $(CH3)_3SPbI3$ ($TMSPbI_3$) perovskite by removing TMS ligands should be semimetallic, rendering it a promising electrode material [6]. Moreover, in view of realizing advanced multi-valued logic devices [7-9], we predicted large room-temperature negative differential resistance (NDR) with high peak-to-valley ratios and current densities can be derived from $PbI_3$-$TMSPbI_3$-$PbI_3$ heterojunction tunneling devices [6]. However, as in the case for solar cell applications [10-13], the toxicity of Pb-based halide perovskites could potentially become the bottleneck for its commercialization.

In this work, adopting a first-principles approach that combines density functional theory (DFT) and nonequilibrium Green's function (NEGF) approaches, we discuss the electronic and quantum transport characteristics of non-toxic TMS germanium triiodide $(CH_3)_3SGeI_3$ ($TMSGeI_3$) and its inorganic framework, germanium triiodide $GeI_3$. We first confirm that the quasi-1D organic-inorganic hybrid halide perovskite $TMSGeI_3$ with a semiconducting character is dynamically stable. Concurrently, we find that its 1D inorganic framework, $GeI_3$, which adopts the face-sharing [$GeI_6$] octahedral geometry, assumes a semi-metallic character like $PbI_3$. Next, considering van der Waals (vdW) bundled quasi-1D heterojunctions in which semiconducting $TMSGeI_3$ channels with sub-5 nm dimensions are sandwiched between semimetallic $GeI_3$ electrodes, we obtain excellent NDR properties characterized by high NDR peak current density (up to ~ 2700 kA.cm$^{-2}$) and peak-to-valley current ratios (PVRs, up to

~44) at low-bias regimes (< 0.8 V). Importantly, we also demonstrate the NDR performances are robustly preserved even after uniaxial strains are applied along the GeI$_3$-TMSGeI$_3$-GeI$_3$ heterojunction column direction, and particularly under a 4 % compressive strain NDR metrics can be further enhanced to the enormous NDR peak current density of 5300 kA·cm$^{-2}$ and PVRs of 87. It will be emphasized that these superb NDR characteristics originate from the quantum hybridization NDR mechanism [6,14].

## 2. METHODS/EXPERIMENTAL

### 2.1 DFT calculations

We performed DFT calculations with the Vienna Ab-initio Simulation Package [15]. The plane-waves were expanded with a kinetic energy cutoff of 500 eV to obtain basis sets with the self-consistency cycle convergence energy criterion of 10$^{-8}$ eV. Atomic structures were optimized using conjugate-gradient approach until the Hellmann–Feynman forces were less than 0.001 eV/Å. The simulations were performed within the Perdew-Burke-Ernzenhof parameterization of generalized gradient approximation revised for solids (PBEsol) [16], which was confirmed to be a suitable exchange-correlation functional for describing TMS-based halide perovskites including their intercolumn vdW interactions [6]. The core and valence electrons were handled by the projector augmented wave method [15]. The $k$-point meshes of 5 x 5 x 8 and 1 x 1 x 8 were employed for unit cells of three-dimensional (3D) TMSGeI$_3$ and 1D GeI$_3$ wire structures, respectively. A vacuum space of more than 15 Å was inserted perpendicular to the periodic 1D structure to avoid interactions with their neighboring images in a periodic boundary condition setup. In order to determine the dynamic stability of TMSGeI$_3$ perovskite and GeI$_3$ inorganic metal-halide nanostructure, we adopted the 2 x 2 x 3 and 1 x 1 x 4 supercells, respectively, and computed the force constant matrices based on small displacement method.

### 2.2 DFT-based NEGF calculations

For the finite-bias non-equilibrium electronic structure calculations, we used the DFT-NEGF method implemented within the TranSIESTA code [17,18]. The surface Green's function $g_s$ were extracted from separate DFT calculations for four unit cells of GeI$_3$ crystals with the 5 × 5 $k_\parallel$-point sampling along the surface $ab$ plane and 10 $k_\perp$-point sampling along the surface-normal charge transport $c$ direction. The transmission functions were then obtained according to

$$T(E;V_b) = Tr[\mathbf{\Gamma}_L(E;V_b)\mathbf{G}(E;V_b)\mathbf{\Gamma}_R(E;V_b)\mathbf{G}^\dagger(E;V_b)], \quad (1)$$

where $\mathbf{G}$ is the retarded Green's function for the channel $C$ and $\mathbf{\Gamma}_{L(R)}$ are the broadening matrices. The current density–bias voltage ($J$-$V_b$) characteristics were calculated by invoking the Landauer-Büttiker formula [19],

$$I(V_b) = \frac{2e}{h}\int_{\mu_L}^{\mu_R} T(E;V_b)[f(E-\mu_R) - f(E-\mu_L)]dE. \quad (2)$$

Analyses on DFT-NEGF calculation output were performed using the Inelastica code and our in-house codes that implement the multi-space constrained-search DFT formalism [14,20,21,22].

## 3. RESULTS AND DISCUSSION

### 3.1. Screening process of metal-halide inorganic frameworks to detect metallicity

A ubiquitous key challenge in hybrid halide perovskite-based device applications is how to eliminate the hazards of Pb exposure [10-13]. Naturally, the approach employing another group 14 metal, Sn or Ge, has been actively explored as a viable option to eliminate the hazardous Pb element. In view of coming up with a non-toxic alternative to the semimetallic 1D PbI$_3$ nanowire and its semiconducting quasi-1D parental perovskite form TMSPbI$_3$, we thus performed the computational screening process (see Fig. 1a, *left*) by adopting the structural template of distorted face-sharing [BX$_6$] octahedral geometry of PbI$_3$ and replacing Pb with Ge or Sn as a cation B in combination with three different types of halogen anions (X: Cl, Br, and I). In Figs. 1b and 1c, we show the atomic structures of nontoxic bulk TMSGeI$_3$ perovskite and its inorganic core GeI$_3$ nanowire, respectively, optimized within the PBEsol [16]. Note that theGeI$_3$ framework composed of connected face-sharing [GeI$_6$] octahedra can be prepared by removing two organic TMS ligands per TMS-GeI$_3$ unit cell. Similar to the experimentally synthesized TMSPbI$_3$ counterpart [23,24], the 3D crystal structure of TMSGeI$_3$ has hexagonal symmetry in the space group *P6$_3$mc* (no. 186) and can be considered as a composite of semi-1D TMSGeI$_3$ columns assembled by vdW interactions.

As the first step of our computational screening pipeline, we considered the BX$_3$ inorganic framework candidates in view of the electronic structure and their dynamical stabilities. The screening results are summarized in the right panel of Fig. 1a, and the optimized atomic structures of $BX_3$ inorganic framework candidates are provided in Fig. 2a. We then observe that Br- or Cl-based metal halides, $GeBr_3$, $GeCl_3$, $SnBr_3$, and $SnCl_3$, as well as $PbBr_3$ and $PbCl_3$, assume semiconducting characters with the B-X bond-length alternation or contracting-expanding Peierls distortions of $[BX_6]$ octahedral cages (Fig. 2a top panel). On the other hand, in the case of I-based metal halides, $GeI_3$, and $SnI_3$, uniform (i.e. without Peierls distortion) Ge-I bond lengths of 2.82 Å and Sn-I bond lengths of 2.93 Å were observed from the optimized $GeI_3$ and $SnI_3$ structures, respectively (Fig. 2a). These bond lengths are comparable to the uniform Pb-I bond lengths of 3.06 Å in the $PbI_3$, nanowire. Then, as hinted by the avoidance of the Peierls distortion and like in the $PbI_3$ case, $GeI_3$ and $SnI_3$ assume semi-metallic characters by preserving a linear dispersion at the Fermi-level (Fig. 2b).

At this point, we provide more explanations on the mechanisms of the emergent semi-metallicity or the avoidance of Peierls distortions in $PbI_3$, $GeI_3$ and $SnI_3$ nanowires. The synthesis of $TMSPbI_3$ demonstrated that, unlike typical amine-based A cations, the sulfur-based $(CH_3)_3S^+$ cation can play a unique role of stabilizing 1D $PbI_3$ frameworks. Then, the removal of TMS ligands from a 1D $TMSBX_3$ nanowire will increase the electron count within the $BX_3$ inorganic framework and form half-filled 1D bands, which typically induce Peierls distortions and open bandgaps. However, as explicitly confirmed above, the quasi-1D nature or circumferential interactions between large I $5p$ lone-pair orbitals can avoid direct interactions between Pb cations or the contracting-expanding distortion of $[BX_3]$ octahedral cages [6]. Namely, the suppression of Peierls distortion in $PbI_3$, $GeI_3$ and $SnI_3$ nanowires can be understood in terms of the quasi-1D character of the $BX_3$ nanowire atomic structure and the large size of I- anions. More detailed discussions can be found in Sec. 2.3 and Supplementary Fig. 9 of Ref. [6].

Next, we explored the dynamic stabilities of $GeI_3$ and $SnI_3$ inorganic frameworks by calculating their phonon spectra. We find that imaginary phonon modes are absent in the phonon band dispersion of $GeI_3$ (Fig. 2c, *left*), which confirms the stability of this 1D semimetallic nanostructure. On the other hand, the phonon spectra of $SnI_3$ displayed significant imaginary modes, which indicates its unstable nature (Fig. 2c, *right*). Their vibrational projected density of states (DOS) revealed that iodine is the major contributor for the low-frequency phonon modes of $BX_3$ frameworks (Fig. 2c). Having identified the inorganic $GeI_3$ column as the promising non-toxic alternative to $PbI_3$, we further confirmed its semi-metallic character by employing Heyd−Scuseria−Ernzerhof (HSE) hybrid functional that corrects the self-interaction error within the local and semi-local DFT exchange-correlation functional [25] (Fig. 2b, middle). In summary, carrying out the screening process, we identified $GeI_3$ as a promising non-toxic 1D semi-metallic material.

Before considering device applications based on $GeI_3$-based heterojunctions, we also discuss the material properties of its parental form, $TMSGeI_3$ perovskite (Fig. 3). In Fig. 3a, we show the calculated electronic band structures of bulk (3D) $TMSGeI_3$ perovskites. At the PBEsol level, we obtain an indirect bandgap of 2.67 eV (Fig. 3a), which is reduced by about 0.43 eV from the bandgap of $TMSPbI_3$ analogues (3.1 eV). We note that this reduced bandgap value is promising in view of photovoltaic applications [5]. Computing the phonon spectrum of $TMSGeI_3$ perovskite (Fig. 3b), we further confirmed the absence of imaginary modes or its high dynamical stability.

## 3.2. Ultrahigh NDR from halide perovskite nanowire junctions and its strain dependence

Adopting the vdW bundled heterojunction nanowires consisting of $GeI_3$-$TMSGeI_3$-$GeI_3$ (Fig. 4a), we next carried out DFT-based NEGF calculations and examined the bias-dependent quantum transport properties. As discussed previously [6], such heterojunction structures could be prepared by selectively peeling off organic TMS ligands from $TMSGeI_3$ through a chemical etching process and exposing stable semimetallic $GeI_3$ columns that can be utilized as electrodes [26]. We previously examined the NDR performance of $PbI_3$-$TMSPbI_3$-$PbI_3$ junctions by varying the length of $TMSPbI_3$ channel length from 3 to 5 unit cells (UCs), and concluded that the five UC (5UC) $TMSPbI_3$ case provides the overall best NDR metrics with the PVR of 17.4 and the peak current density of ~ 921 kA·cm$^{-2}$. We thus adopted a similarly-dimensioned 5UC $TMSGeI_3$ channel and present the calculated $J$–$V_b$ characteristics in Fig. 4b (black solid line). With the sub-5 nm long channel, we obtain excellent NDR performances characterized by a high PVR up to 44.3 and a very high peak current density reaching ~ 2741 kA·cm$^{-2}$ achieved at low-bias voltage regimes (< 0.8 V). Particularly, compared to the NDR device metrics of the reference $PbI_3$-5UC $TMSPbI_3$-$PbI_3$ counterpart (gray dashed line in Fig. 4b), we find that those from the $GeI_3$-$TMSGeI_3$-$GeI_3$ junction are far superior except that the NDR peak and valley appear at slightly higher bias voltage values of $V_b$ = 0.5 V and 0.8 V, respectively.

To explain the mechanisms of the appearance of NDRs in GeI$_3$-TMSGeI$_3$-GeI$_3$ junctions, we show in Fig. 4c the development of projected local electronic DOS across a GeI$_3$-TMSGeI$_3$-GeI$_3$ junction with increasing $V_b$ values. The first notable feature is that at $V_b$ = 0.0 V (Fig. 4c left panel) the hole Schottky barrier height (SBH) at the TMSGeI$_3$-GeI$_3$ interface is only ~0.25 eV, which apparently originates from the fact that the same GeI$_3$ inorganic framework is shared throughout the GeI$_3$-TMSGeI$_3$-GeI$_3$ junction. The marginal hole SBH then allows the appearance of ample metal induced gap states (MIGS) spatially within the TMSGeI$_3$ channel region and energetically between the TMSGeI$_3$ valence band maximum (dotted lines) and the GeI$_3$ Fermi levels (solid lines). Different from conventional MIGS [27,28], they are quantum-hybridized states entangling two GeI$_3$ electrode states and the special electrode-channel-electrode quantum-hybridized character can be confirmed by observing their response to finite bias voltages and corresponding transmissions. Specifically, upon increasing the applied bias, we find that until $V_b$ = 0.5 V (NDR peak; Fig. 4c middle panel) that corresponds to twice of the hole SBH (0.25 eV) the MIGS bound by quasi-Fermi levels (dotted lines) tilt symmetrically [22], maintaining the hybridization across the TMSGeI$_3$ channel and producing large transmission values. However, upon further increasing the bias to $V_b$ = 0.8 V (NDR valley; Fig. 4c right panel), we observe that the spatial hybridization becomes abruptly broken and MIGS are distributed into an asymmetric form (MIGS accumulated near the left GeI$_3$ electrode) with negligible transmission values. This quantum-hybridization NDR mechanism will be once more explained below based on molecular projected Hamiltonian eigenstates (see Fig. 5d).

The differences between the quantum-hybridization NDR performance of the GeI$_3$-TMSGeI$_3$-GeI$_3$ junction and that of the PbI$_3$-TMSPbI$_3$-PbI$_3$ counterpart [6,14] can be then understood in terms of the differences in SBHs (~ 0.25 eV at the GeI$_3$-TMSGeI$_3$ interface vs. ~ 0.15 eV at the PbI$_3$-TMSPbI$_3$ interface) and channel lengths (36.6 Å of the 5UC TMSGeI$_3$ vs. 39.8 Å of the 5UC TMSPbI$_3$). Specifically, compared with the PbI$_3$-TMSPbI$_3$-PbI$_3$ case, the shorter channel length (larger SBH) of the GeI$_3$-TMSGeI$_3$-GeI$_3$ junction results in the increased NDR peak current density (bias voltage position). The shift of the NDR peak position to a higher bias regime will result in a similar upshift of the NDR valley position. This will then allow a more dramatic collapse of quantum-hybridized states, which should translate into the reduction of the NDR valley current density or the enhancement of NDR PVR.

Finally, in view of wearable and flexible electronics applications, we applied uniaxial strain along the *c*-axis at constant volume by compressing and stretching the GeI$_3$-TMSGeI$_3$-GeI$_3$ junction and repeated the room temperature transport calculations. Figures 5a and b show the *J–V$_b$* characteristics of the GeI$_3$-TMSGeI$_3$-GeI$_3$ junctions with 4% compressive (red circle) and 4% tensile (blue triangle) strain applied and the corresponding transmission spectra, respectively. In Fig. 5a, we can confirm the robustness of NDR signals at low-bias operating conditions (< 0.8 V) regardless of the applied strain. While the NDR peaks appear at more or less similar $V_b$ value of ~ 0.5 V, the NDR valleys are shifted to higher $V_b$ values with increasing compressive strain. The subsequent analysis of the electronic structures of bundled GeI$_3$ and TMSGeI$_3$ with compressive and tensile strain along the *c*-axis clarifies that the semimetallicity of GeI$_3$ is robustly preserved within ±4% uniaxial strain conditions (Fig. 5c). Overall, we found that the application of a compressive strain further leads to the enhancement of the NDR performance: Compared to the unstrained junction, the PVR value significantly increases from 44.3 to 87.1 at 4 % compression. Moreover, with the 4 % compression, the peak current density is significantly enhanced from 2741 kA·cm$^{-2}$ to 5365 kA·cm$^{-2}$ (Fig. 5a). On the other hand, the decrease in the peak current density upon stretching was obtained as shown in Fig. 5a. These variations in the peak current density can be understood in terms of the change in the coupling strength between electrode and channel states [19]. Via shortening the distance between GeI$_3$ electrodes, the hybridization of TMSGeI$_3$-GeI$_3$ interface states and accordingly their spatial extensions are substantially increased. This can be directly visualized through the molecular projected Hamiltonian eigenstates [20-22] that contribute most strongly to quantum transport (orange left triangle in Fig. 5b): Compared to the unstrained case, as shown in the top panel of Fig. 5d, the compressive strain or the shortened TMSGeI$_3$-GeI$_3$ interfacial bond length results in strong delocalization of interfacial states into the channel region. On the other hand, under the tensile strain, the extended TMSGeI$_3$-GeI$_3$ interfacial bond length should cause the weakening of their coupling strength and decrease the peak current density.

## 4. CONCLUSION

In summary, carrying out combined DFT and NEGF calculations, we explored structural, electronic, and charge transport properties of the lead-free non-toxic hybrid halide perovskite TMSGeI$_3$ nanowires, their GeI$_3$ inorganic frameworks, and GeI$_3$-TMSGeI$_3$-GeI$_3$ junctions. Through a computational screening process, we first identified that the 1D GeI$_3$ inorganic framework that adapts a face-sharing [GeI$_6$] octahedral geometry exhibits a metallic behavior without Peierls distortion and is dynamically stable. Concurrently, we confirmed the semiconducting character of the quasi-1D parental TMSGeI$_3$ perovskite nanowires as well as its dynamical stability. Next, adopting the van der Waals bundled nanowire heterojunction structures in which TMSGeI$_3$ channels with sub-5 nm dimensions are sandwiched between GeI$_3$ electrodes, we predicted that excellent NDR characteristics can be obtained. Characterized by drastically high peak current density (~ 2741 kA·cm$^{-2}$) and room-temperature resistive switching ratio (PVR ≈ 44.3), we emphasized that these NDR metrics emerge from the quantum hybridization NDR mechanism. Finally, in view of flexible electronics applications, we confirmed that the NDR performances are robustly preserved under uniaxial tensile and compressive strains and particularly the NDR peak current density and PVR can be further enhanced to 5365 kA·cm$^{-2}$ and 87.1, respectively, under 4 % compressive strain. Our work demonstrates the significant potential of low-dimensional hybrid halide perovskites for the realization of beyond-CMOS and wearable flexible electronic devices.

## ACKNOWLEDGMENTS

This work was supported by the Samsung Research Funding & Incubation Center of Samsung Electronics (No. SRFC-TA2003-01). Computational resources were provided by KISTI Supercomputing Center (KSC-2018-C2-0032).

## AUTHOR CONTRIBUTIONS

YHK oversaw the project, and JL and MEK carried out calculations. All authors analyzed the computational results and co-wrote the manuscript. All authors read and approved the final manuscript.

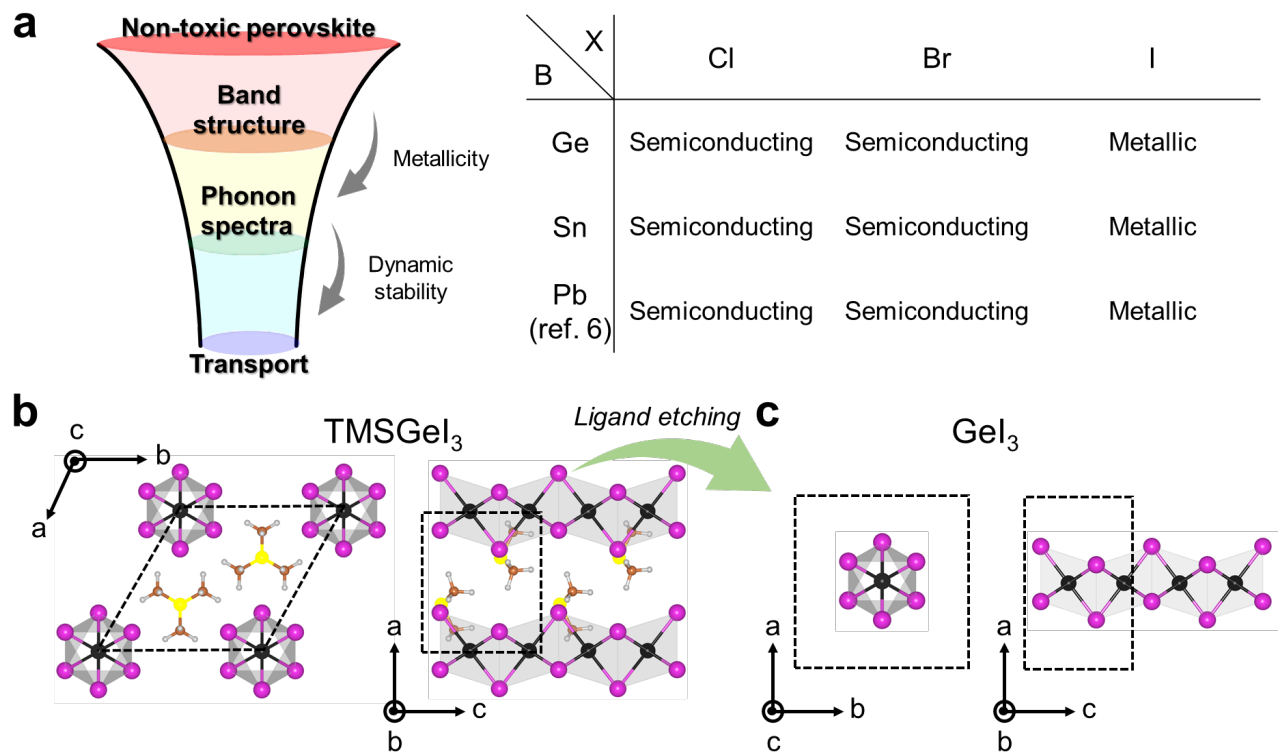

**Fig. 1 a** Schematic of a tiered "funnel" screening pipeline (*left*) utilized to discover low-dimensional hybrid halide perovskites for NDR devices. The screening results are summarized in the table (*right*). **b** Crystal structures of quasi-1D bulk TMSGeI$_3$ and **c** 1D GeI$_3$ inorganic core-only face-sharing octahedral framework. The black dotted boxes indicate the unit cells for each case.

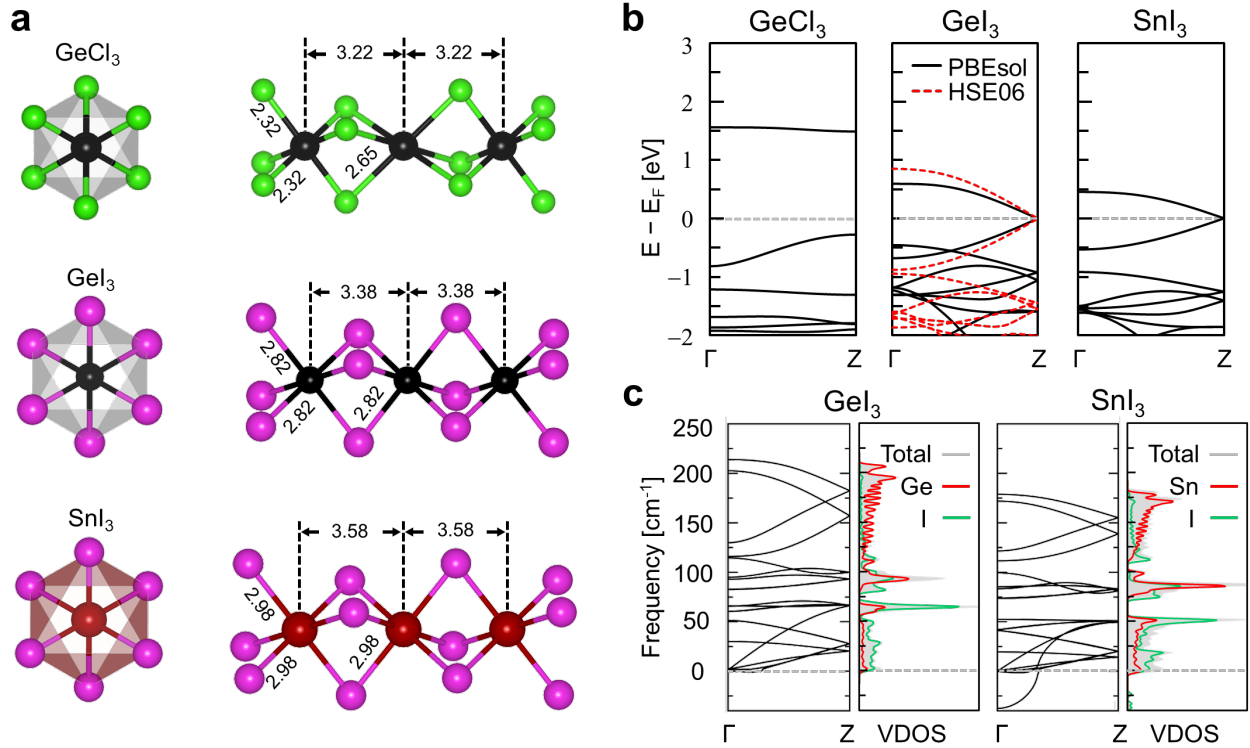

**Fig. 2 a** The atomic geometries of 1D face-sharing octahedral frameworks of Peierls-distorted $GeCl_3$, and Peierls distortion-avoiding $GeI_3$, and $SnI_3$ inorganic nanowires. **b** The electronic band spectra showing the semiconducting character of $GeCl_3$ and semimetallic nature of $GeI_3$ and $SnI_3$ nanowires. For $GeI_3$, in addition to the PBEsol band structure (black solid lines), HSE data are presented (red dotted lines). **c** Phonon band spectra (left panels) and corresponding vibrational DOS (VDOS; right panels) of $GeI_3$ and $SnI_3$ nanowires. For the vibrational DOS, we show the projections to Ge or Sn (red lines), and I (green lines).

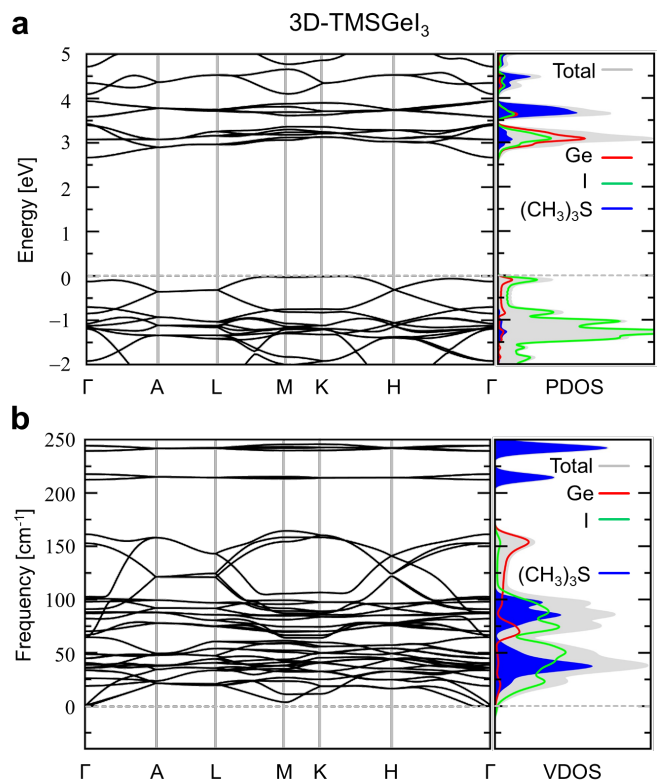

**Fig. 3 a** Electronic band structures and **b** phonon band spectra obtained from the quasi-1D TMSGeI$_3$ perovskite bulk structure calculated within PBEsol. For both electronic and vibrational DOS, we show the projections to Ge (red lines), I (green lines), and TMS (blue lines). In **b**, for clarity, high-frequency phonon modes (>250 cm$^{-1}$) originating from (CH$_3$)$_3$S$^+$ ligands were omitted.

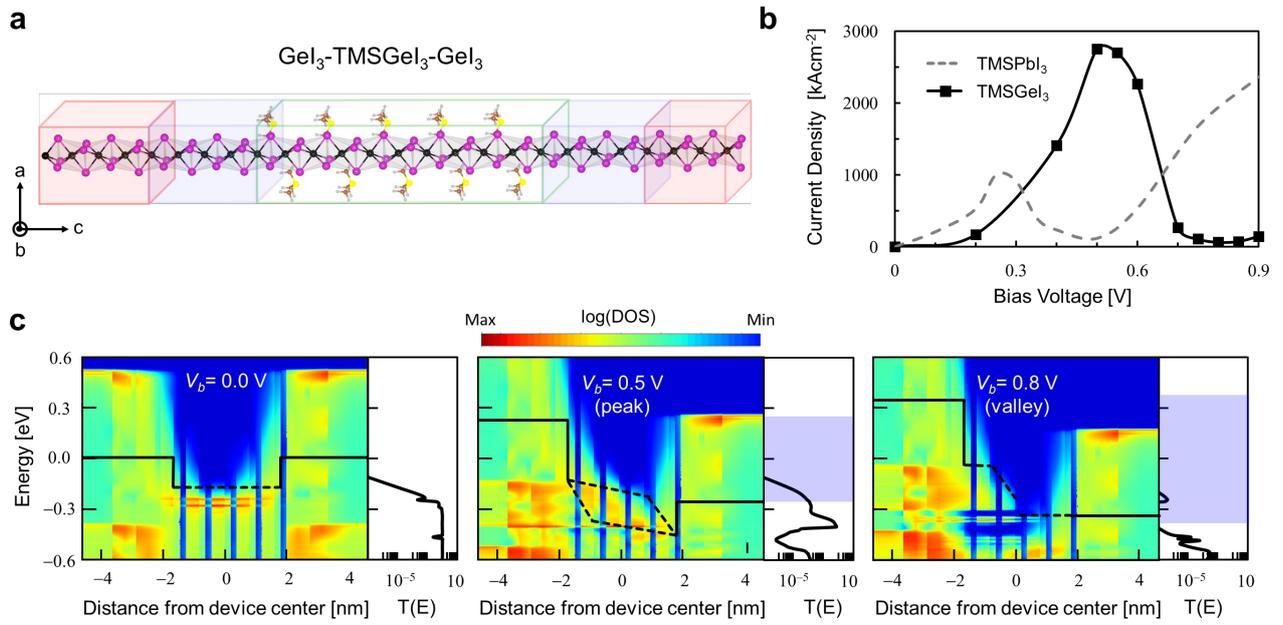

**Fig. 4 a** The optimized atomic structure of van der Waals bundled GeI$_3$-TMSGeI$_3$-GeI$_3$ nanowire junctions based on the 5UC TMSGeI$_3$ channel. Red and blue boxes indicate the electrode regions that are replaced by separate semi-infinite electrode models and retained as scattering regions, respectively, within NEGF quantum transport calculations. **b** The $J$-$V_b$ characteristics of the GeI$_3$-5UC TMSGeI$_3$-GeI$_3$ and PbI$_3$-5UC TMSPbI$_3$-PbI$_3$ junctions. **c** Valance band maxima-region projected local DOS (left panels) and transmission spectra (right panels) of the GeI$_3$-5UC TMSGeI$_3$-GeI$_3$ junction at $V_b$ = 0 V (left) 0.5 V (middle), and 0.8 V (right), respectively. Conduction band minimum-region data are not shown for clarity. Solid and dotted lines indicate the Fermi levels in the GeI$_3$ electrodes and the quasi-Fermi levels in the TMSGeI$_3$ channel, respectively. Blue shaded boxes in **b** and **c** indicate the applied voltage windows.

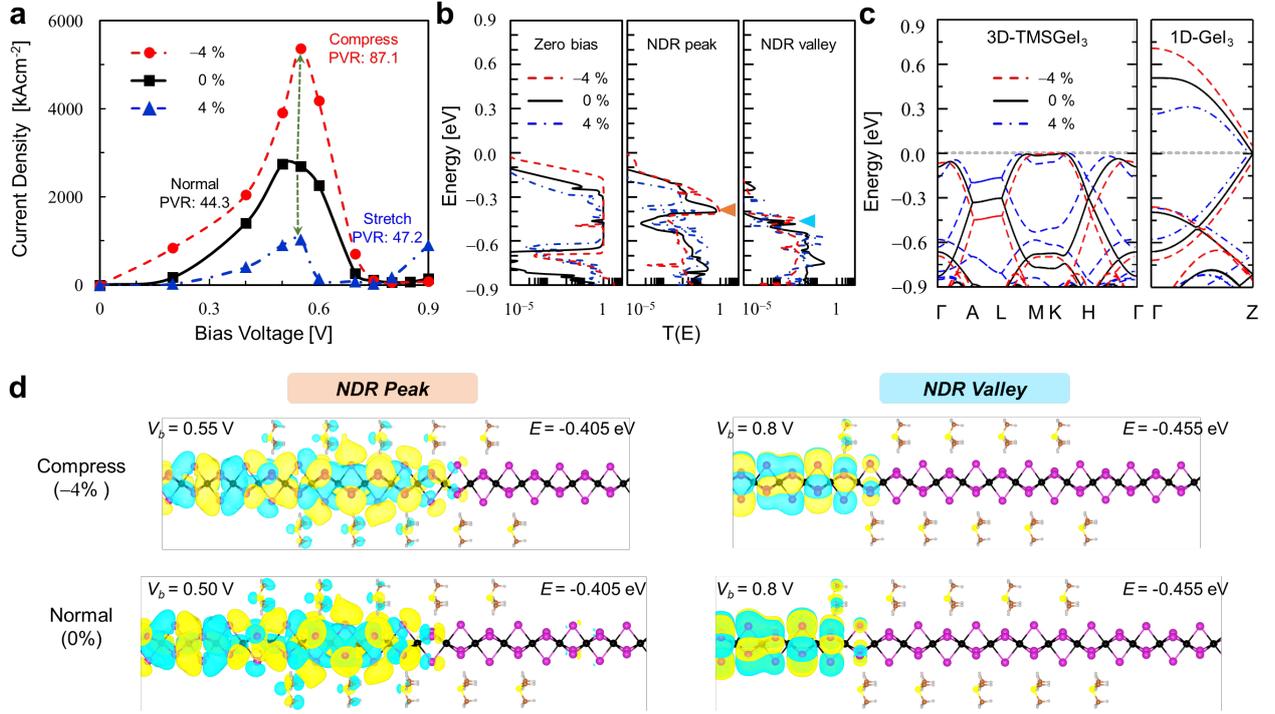

**Fig. 5 a** The NDR $J$-$V_b$ characteristics of the van der Waals bundled GeI$_3$-5UC TMSGeI$_3$-GeI$_3$ junction under +4 % (compressive), 0 %, and −4 % (tensile) uniaxial strains. **b** Electronic transmission spectra of the unstrained as well as strained device at different $V_b$ values. Orange and cyan left triangle indicate the points that contribute most strongly to quantum tunneling at NDR peak and valley, respectively. **c** Electronic band structures of 1D GeI$_3$ columns and quasi-1D TMSGeI$_3$ at +4 %, 0 %, and 4 % uniaxial strains. **d** Molecular projected Hamiltonian states for the NDR device with compressive strain (*top*) and without strain (*bottom*) at the NDR peak (*left*) and valley (*right*) positions, respectively. The isosurface level is 3×10$^{-3}$ Å$^{-3}$.

11